# Effect of Te doping in GeSe parent thick film by experimental *in situ* temperature-dependent structural investigation.

P. Armand*, R. Escalier, G. Silly, J. Lizion, A. Piarristeguy

ICGM, Univ. Montpellier, CNRS, ENSCM, Montpellier, France

*Corresponding author: pascale.armand@umontpellier.fr

| | | |
|---|---|---|
| ORCID: | P. Armand | 0000-0001-8921-5427 |
| | A. Piarristeguy | 0000-0002-8922-4566 |
| | G. Silly | 0000-0001-8212-2359 |
| | J. Lizion | 0000-0001-9433-5394 |
| | R. Escalier | 0000-0003-4179-804X |

**Abstract:**

As-deposited and unencapsulated GeSe$_{1-x}$Te$_x$ ($x$ = 0, 0.25) 3-µm-thick amorphous films on Si(001) were obtained via the co-evaporation technique to study the effect of selenium (Se) substitution for tellurium (Te) on the GeSe parent structure in the function of the heating temperature. *In situ,* grazing-incidence X-ray scattering (XRS), and fluorescence X-ray Absorption Near Edge Structure (XANES) data were collected in isochronal annealing conditions under nitrogen flow. The results show that the onset temperature of crystallization $T_c$ and the crystallized phase symmetry are susceptible to the Te 25 at. % doping. **Furthermore, Ge and Se K-edge XANES analyses reveal significant alterations in the local atomic environments surrounding Ge and Se atoms during the transition from the amorphous to the crystalline state. These modifications are accompanied by an observable increase in local structural disorder upon substitution with Te atoms.**

1. **Introduction**

Binary IV-VI chalcogenides are well-studied materials as they present many distinctive electrical, thermal, and optical characteristics that have the potential to be beneficial for a variety of applications [1−3]. GeSe and GeTe, end-formers of the pseudo-binary GeSe-GeTe in the Ge-Te-Se ternary system, present high structural and physical property differences. However, Se and Te are from the same periodic table group.

The semiconductor α-GeTe exhibits, at room temperature, a rhombohedral structure (space group of *R3m*, N°160 [4, 5]) and is recognized as a prominent phase-change material (PCM) [6−8]. It undergoes a fast and reversible transition between its amorphous and crystalline states. It showcases significant optical and electrical differences crucial for data storage, contrasting between set and reset bits. The phase change property of GeTe-based materials would be due to a unique bonding mechanism called "metavalent bonding" (MVB) forming their rhombohedral structure and by the coordination number modification of the Ge atoms between the amorphous and the crystallized states [6−8]. MVB is a novel type of bonding characterized by the competition between electron delocalization as in metallic bonding and electron localization as in covalent bonding [6−8].

α-GeTe is a non-stoichiometric material with vacancies at Ge sites in the crystal lattice that lead to crystallographic defects and dangling bonds and can lead to the extrusion of a minor parasitic phase of elemental Ge [9−11]. Moreover, Te chalcogen is more metallic compared to the chalcogen Se, then, the formation of GeTe glasses needs ultra-fast quenching [12].

On the other hand, group-IV monochalcogenide α-GeSe adopts, under ambient conditions, an orthorhombic structure with *Pnma* (N° 62) space group [13, 14]. Each unit cell consists of two GeSe sheets oriented in the (0, 0, 1) direction with weak *van der Waals* interactions between the layers. It is particularly attractive for near-infrared (NIR) photodetectors, electronic tunneling devices, and photovoltaic absorber material [15, 16]. This covalent and stoichiometric material presents several advantages: low dangling bond issue, high chemical stability, and low surface defect density.

Pristine α-GeSe is not known as a phase change material. The $Ge_4Se_3Te$ compound corresponds to the doping of GeSe composition by replacing 25 % of the selenium atoms of the unit cell with the tellurium ones ($GeSe_{0.75}Te_{0.25}$). This material exists in two crystalline structures at ambient conditions. First, in a hexagonal structure (*$P6_3mc$* space group), when the system is in thermodynamic equilibrium, stable only below ~400 °C [17–19]. This $Ge_4Se_3Te$ hexagonal structure does not present a portfolio of properties compatible with metavalent-bonded materials but with ordinary covalent bonding. Then, a metastable rhombohedral α-GeTe-like structure (*R3m*) is found when the system is out of equilibrium [20]. The existence of a stable rhombohedral structure with $Ge_4Se_3Te$ composition has never been reported for a bulked solid solution under thermodynamic equilibrium [21–23].

The out-of-equilibrium conditions are encountered in the PCM memories, thus, the $Ge_4Se_3Te$ composition in its rhombohedral structure can be considered a good PCM compromise, and consequently, the transition from the amorphous to the crystallized states needs to be studied.

This paper deals with the structural investigations of two compositions, the GeSe parent material and $GeSe_{0.75}Te_{0.25}$, the Te-doped compound, deposited on Si substrates as thick amorphous films of 3 μm, by the thermal co-evaporation technique. *In situ*, temperature-dependent X-ray scattering experiments coupled with fluorescence measurements at both Se and Ge K-edges followed the amorphous local order, the transition temperature, and the nature of the crystallized phase. This work aims to determine i) if modifications of the order at short and medium distances exist between the amorphous and crystallized state, and ii) if structural evolutions exist between the undoped and the Te-doped composition.

## 2. Experimental section

*2.1 Samples synthesis*

Films of GeSe$_{1-x}$Te$_x$ (with $x$ = 0, 0.25) nominal compositions measuring 3 micrometers in thickness were deposited on a Si (001) substrate through the thermal co-evaporation method, utilizing high-purity elements (Ge pieces sourced from Goodfellow, Te pieces, and Se granules from ChemPur, each boasting a purity level of 99.999%). Depositions were carried out using a PLASSYS MEB 500 device, with the substrate undergoing rotation to ensure uniform film thickness. More details on the experimental conditions can be found **elsewhere** [20].

**The as-deposited amorphous state of these films was confirmed by X-ray scattering at room temperature; the experiments were carried out on a Panalytical X'Pert diffractometer using the Cu Kα radiation.** Notably, the thick films were characterized without capping layers, as previous research has demonstrated that oxidation does not significantly impact the characteristic temperatures of GeTe films exceeding 50 nm in thickness [20, 24].

*2.2 Characterizations*

Simultaneous collection of temperature-dependent grazing incidence fluorescence and X-ray scattering (XRS) patterns was conducted *in situ* at the DiffAbs beamline within the SOLEIL Synchrotron facility (www.synchrotron-soleil.fr) located in Saint-Aubin, France. **This made it possible to study the same sample under identical conditions using two complementary methods.** A prior publication references comprehensive information regarding the experimental setup and furnace specifications [20]. The furnace was mounted on a six-circle diffractometer, and the grazing incident angle ω was fixed to 6°. The X-ray photon energy was set to 11.015 keV (λ = 1.1256 Å) using a double-crystal Si (111) monochromator (with an energy resolution $\Delta E/E = 10^{-4}$). **To acquire the X-ray Absorption Near Edge Structure (XANES) spectra, energetic ranges corresponding to the characteristic K-edge absorption lines of both Ge and Se elements were precisely delineated.** To optimize acquisition efficiency, fluorescence-type XANES spectra were acquired at the Ge K-edge and Se K-edge with a variable energy step. The energy of the K threshold of Te atoms (31.814 keV) is not accessible on the DiffAbs beamline whose energy range is 3 - 23 keV. Elastic X-ray scattered signals were detected utilizing an XPAD-S140 two-dimensional hybrid pixel detector [25, 26].

For the GeSe parent composition, the *in-situ* temperature-dependent data acquisition loop, repeated from 25 to 370 °C, involved 3 iterations of XRD patterns, each lasting 30 seconds with a 2θ step of 0.0115°, followed by one Ge K-edge XANES spectrum recorded at a rate of 2 seconds per point, totaling 8 minutes for a full spectrum and then, one XANES spectrum at the Se K-edge (2 seconds per point, 8 minutes for a full spectrum). Throughout the temperature ramps, experimental data were continuously logged. Initially, the temperature ramp was set at 20 °C/min from 25 to 250 °C (amorphous state), then adjusted to 0.2 °C/min up to 370 °C. Before each sequence, automatic realignment of the sample surface was performed. A similar procedure was adopted for the substituted film $GeSe_{0.75}Te_{0.25}$. The temperature ramp followed the pattern of 20 °C/min from 25 to 200 °C (amorphous state), then 0.2 °C/min up to 376 °C.

The XANES spectra were normalized in absorbance through the following steps:

1. The raw data in the pre-edge region (90 to 30 eV before the edge step $E_0$) were fitted with a first-order polynomial. This polynomial was extrapolated through the post-edge region and subtracted as background absorption.

2. The spectra were normalized for atomic absorption using a second-order polynomial fitted to the spectral region from 15 to 120 eV after the edge.

3. The pre-edge line was subtracted from the entire spectra, and the resulting spectra were divided by the absorption edge step.

The edge step was determined by taking the difference between the pre-edge and normalization lines at $E_0$, with $E_0$ defined as the maximum of the first derivative curve.

**3. Results**

*3.1 GeSe parent composition*

*3.1.1 Temperature-dependent synchrotron X-ray scattering patterns*

The grazing incidence X-ray scattering patterns were recorded *in situ* with a continuous temperature rise on the uncapped parent film of GeSe nominal composition using a monochromatic beam (wavelength λ of 1.1256 Å), Figure 1. Figure 1(a) **shows** a comprehensive 2D overview of the structural evolution of the GeSe film across the entire temperature range studied, without any thermal discontinuities. This figure precisely delineates different thermal zones (I, II, III), corresponding to various phase states. This representation of the *in-situ* X-ray scattering (XRS) data facilitates easy comparison of the thermal domains where amorphous and crystalline states coexist, which, in turn, indicate the glass crystallization rate and phase stability. Additionally, this plot offers a visual means to identify the formation of metastable phases that may exist within a narrow temperature range of a few degrees Celsius. In contrast, the X-ray diffractograms at specific temperatures (Figures 1(b)) allow for detailed structural analysis at key temperatures. The combination of these two types of data—broad temperature range overviews and specific temperature

diffractograms—provides a thorough understanding of the structural properties and glass crystallization behavior as a function of temperature.

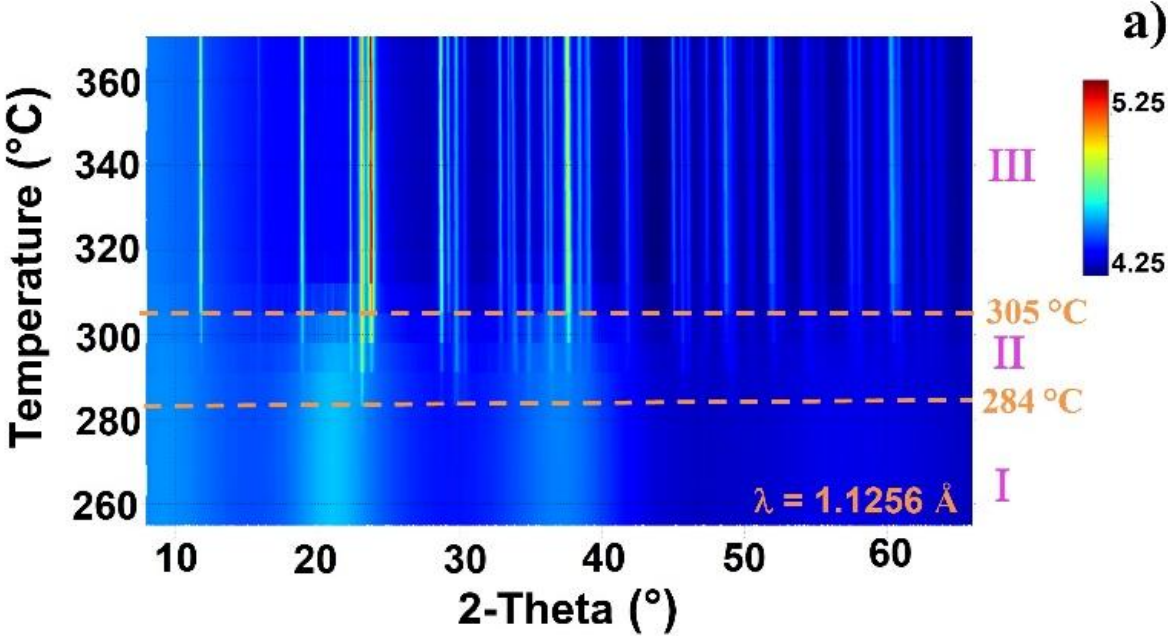

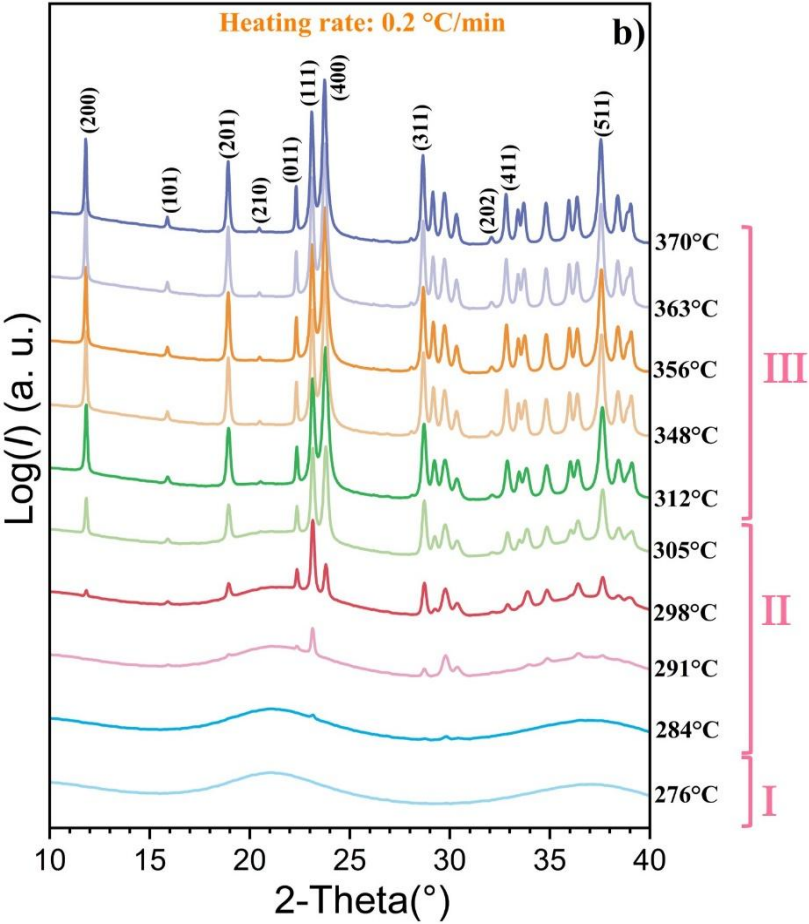

**Figure 1:** a) *In-situ* temperature-dependent synchrotron XRS patterns registered in isochronal annealing conditions on a GeSe 3-μm-thick film deposited on a Si(001) substrate (log scale in color), b) XRS patterns at selected temperatures. The spectra have been stacked for clarity. Only the main reflections are indexed on the graph. **The wavelength of X-ray is 1.1256 Å**. The heating rate was 0.2 °C/min.

From room temperature up to ~284 °C, zone I, the film is amorphous as only large rings are visible. Using XRS, we defined the onset crystallization temperature $T_c$ as the temperature at which diffraction peaks emerge. $T_c$ is at 284 °C, with our experimental conditions. For 284 °C ≤ T ≤ 305 °C, there is a coexistence of the amorphous (rings) and the crystalline (diffraction peaks) states, zone II, which proportion evolutes with the heating. Finally, only diffracted peaks are visible for T > 305 °C, zone III. These diffracted peaks are all associated with the stoichiometric α-GeSe material following the ICDD card n° 96-900-8784 [27] which presents an orthorhombic structure with *Pnma* (N° 62) space group [13, 14].

α-GeSe crystallizes in a highly anisotropic layered orthorhombic structure (distorted NaCl-type) with a Ge 4*s* lone electron pair pointing to the interlayer spacing. The unit cell contains four chemical units forming double-layer slabs of Ge–Se in a chair configuration along the *a*-axis. The layers are separated by van der Waals forces, making α-GeSe easily cleavable along its *bc* plane. Each Ge atom has six unlike atoms as first neighbors: 3*Ge–Se covalent bonds (one shorter (~2.57 Å) and two longer (~2.58 Å)) forming $GeSe_3$ triangular pyramidal structural units, represented by green color lines in Figure 2, and 3*Ge…Se interactions (bicolor blue and pink lines in Figure 2) with more distant neighbors (2*~3.31 Å + 1*~3.37 Å) [13, 14]. Each Se atom is bonded to three Ge as its nearest neighbors and three Ge as its second nearest neighbors.

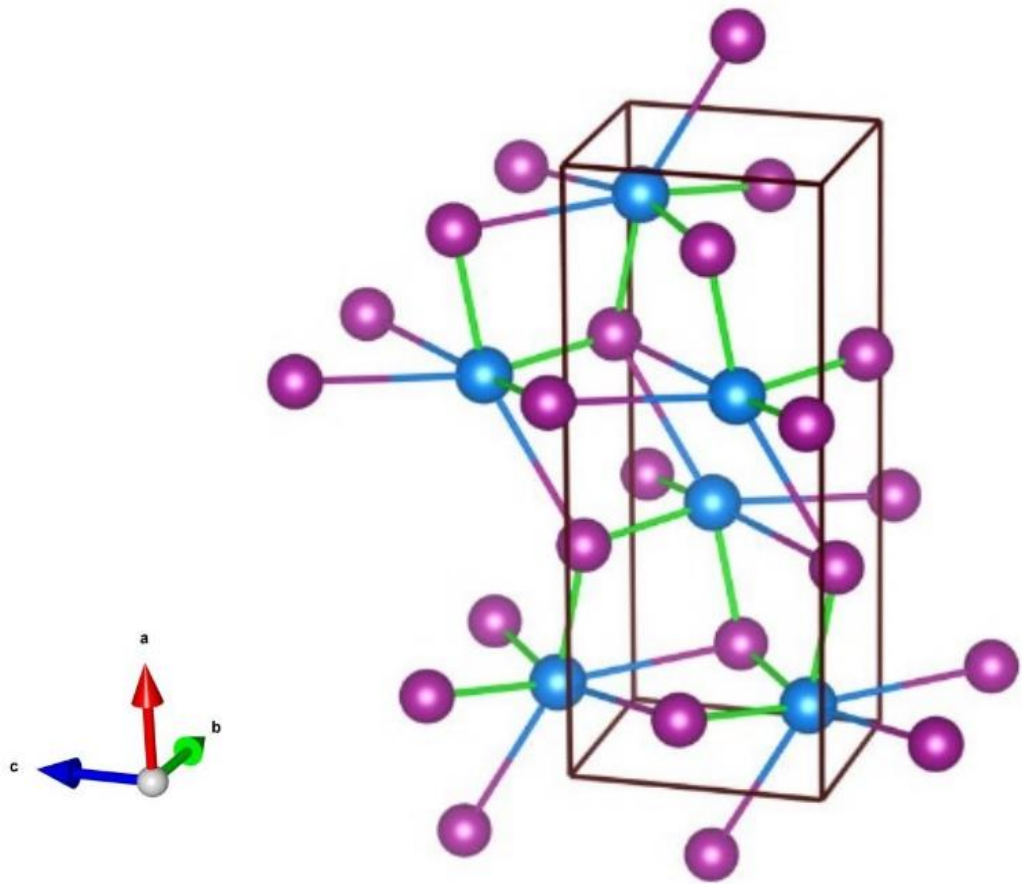

**Figure 2:** Bilayer crystal structure of α-GeSe (*Pnma*). The black line is the unit cell with two GeSe double-layers separated by van der Waals forces. Ge in blue and Se in pink balls. The 3 first-nearest Ge-Se bonds are shown with green lines and the 3 next-nearest nonbonding lengths are with bicolor lines. For clarity, the first Se shell is not shown.

There is no trace of crystallized $GeSe_2$ and/or elemental Ge/Se and no trace of XRD peaks of $GeO_2$ which may indicate the lack of oxidation of the uncapped film's surface at high temperatures. We also noticed that the 100 % intensity corresponds, here, to the (400) diffraction peak (2θ = 23.76°) while it is for the (111) peak in the ICDD card n° 96-900-8784 for a bulk specimen. Thus, α-GeSe crystallizes in a preferential crystallographic direction belonging to the (*h*00) planes (*h* pair) as already mentioned [28].

The numerical results of lattice determinations obtained via the le Bail-type refinements of the powder diffractograms are listed in Table I as a function of the heating temperature. The corresponding unit cell volumes *V* are also given. Room-temperature lattice parameters from a bulk GeSe sample [28] are also reported for comparison.

**Table I:** The temperature dependence of the lattice parameters and unit cell volume *V* of the orthorhombic α-GeSe thick film with *Pnma* structure.

| T | *a* | *b* | *c* | *V* |
|---|---|---|---|---|
| °C | Å | Å | Å | Å$^3$ |
| 25 [28] | 10.82$_2$ | 3.83$_5$ | 4.39$_0$ | 182.26 |
| 312 | 10.91$_5$ | 3.87$_0$ | 4.38$_5$ | 185.24 |
| 320 | 10.91$_6$ | 3.87$_1$ | 4.38$_5$ | 185.26 |
| 327 | 10.91$_9$ | 3.87$_2$ | 4.38$_5$ | 185.38 |
| 334 | 10.91$_9$ | 3.87$_3$ | 4.38$_4$ | 185.40 |
| 341 | 10.92$_5$ | 3.87$_5$ | 4.38$_6$ | 185.66 |
| 348 | 10.92$_7$ | 3.87$_6$ | 4.38$_5$ | 185.72 |
| 355 | 10.93$_0$ | 3.87$_7$ | 4.38$_5$ | 185.84 |
| 363 | 10.93$_4$ | 3.87$_9$ | 4.38$_5$ | 186.00 |
| 370 | 10.93$_8$ | 3.88$_1$ | 4.38$_6$ | 186.17 |

The graphical representation shows that the positive thermal expansion is linear along the two *a* and *b* crystallographic axes, Figure 3(a). The cell volume *V* increases also linearly with the heating temperature, Figure 3(b). The results of the linear least-square extrapolation of the experimental data in function of the temperature are given in the graphs, Figure 3(a) for *a, b,* and *c* experimental lattice parameters, and Figure 3(b) for the unit cell's volume *V*. The slope of the fit is considerably smaller for the *c*-axis than for the *a*- and *b*-axis in this temperature interval (312–370 °C); the *c*-axis has zero thermal expansion.

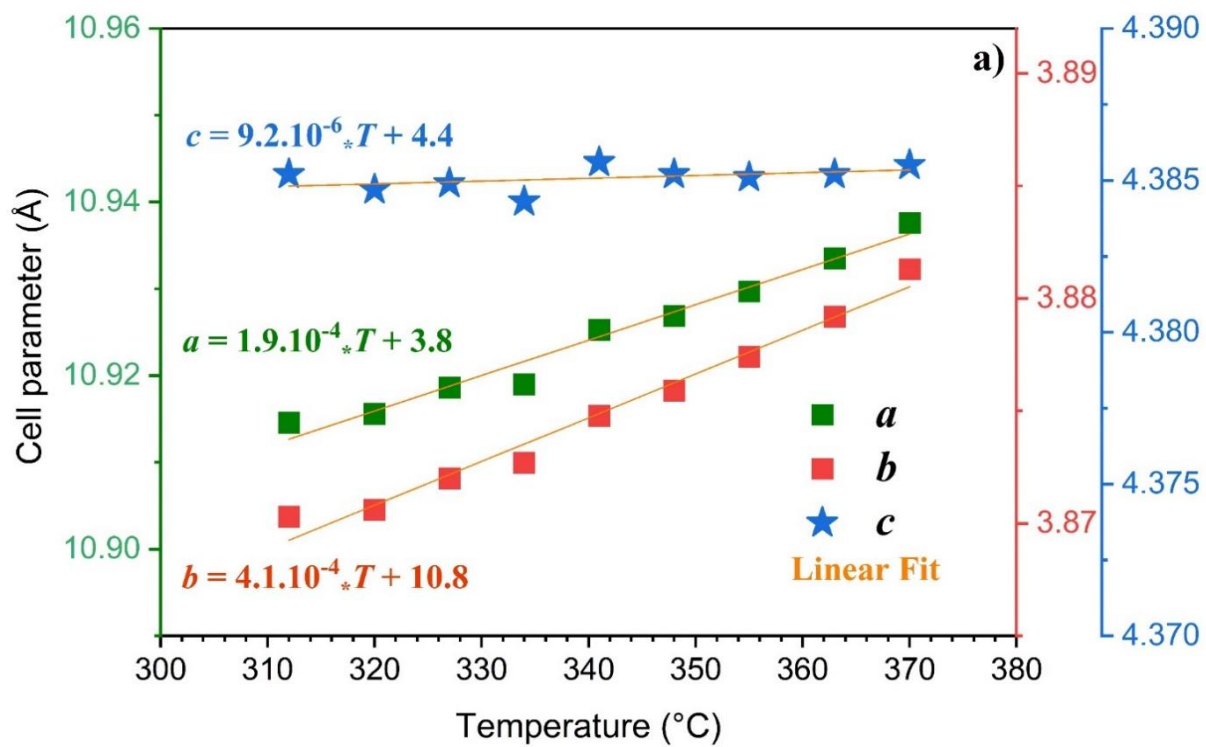

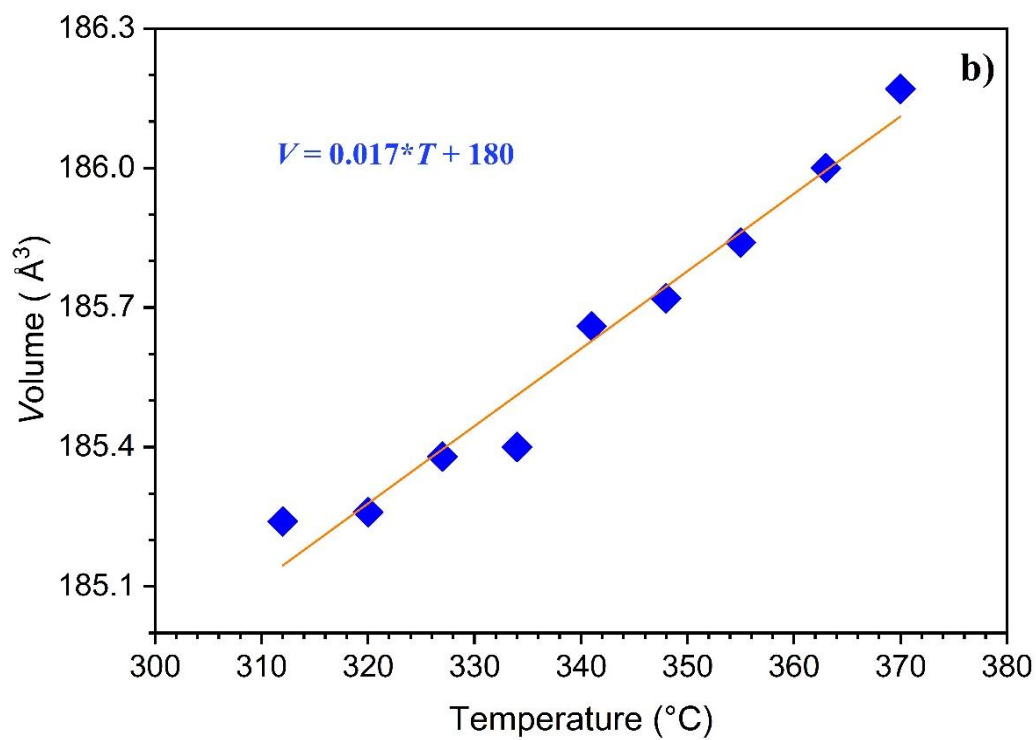

**Figure 3:** Temperature range (310 to 370 °C, zone III) showing the evolution of (a) the lattice parameters *a, b,* and *c,* and (b) the unit cell's volume *V* of the crystalline phase α-GeSe with an orthorhombic structure. Solid lines are linear fitted curves of the experimental data. Error bars are smaller than the symbols. Heating rate **is** 0.2 °C/min.

*3.1.2 Temperature-dependent XANES spectra*

Figure 4(a) presents the fluorescence-detected XANES spectra of the as-deposited GeSe parent thick film registered in continuous heating from room temperature up to 370 °C. Ge K-edge occurs at 11104 eV, with a natural core hole width of 1.96 eV [29] responsible for the low energy resolution of the XANES features. In a solid, the resonant XANES feature (named A in Figure 4) close to the energy of the core excitation level ($E_0$) is related to the electronic structure with dipole transitions from the excited core level to vacant states above the Fermi level, usually *p* or *d* orbital levels. The more vacant states there are in these orbitals, the higher the amplitude of the white line will be. Ge K-edge XANES sharp rise (absorption edge) corresponds to the allowed electronic transition from the Ge 1*s* ground state to empty *p*-like states in the conduction band. The XANES features reflect mostly the 4*p* electronic structure through the partial density of unoccupied states around Ge atoms.

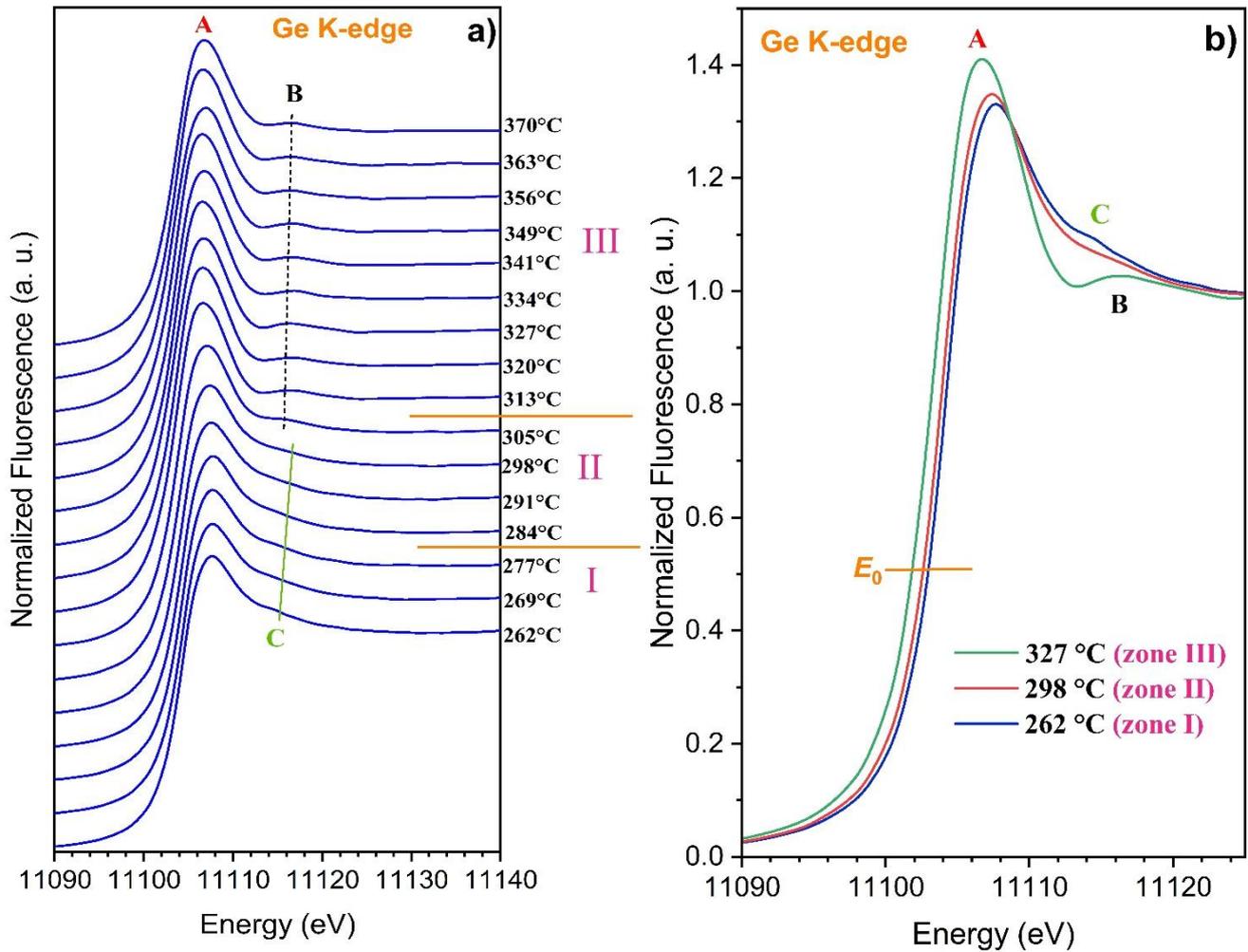

**Figure 4:** a) Evolution with the heating temperature registered in isochronal annealing conditions of the near-edge features in the fluorescence spectra registered at the Ge K-edge of the as-deposited amorphous GeSe thick film. The spectra have been stacked for clarity. b) Fluorescence spectra at selected temperatures. The heating rate was 0.2 °C/min.

Three distinctive experimental XANES spectral shapes exist in the function of the heating temperature. They exhibit XANES resonances labeled from the edge, A, B, and C. The first corresponds to the amorphous state of GeSe (zone I in Figures 1 and 4(b)). The second (zone II), combines the XANES shape of the amorphous and the crystalline states. The third distinctive spectral shape (zone III) is the fingerprint of the α-GeSe framework with an orthorhombic structure. In α-GeSe, Ge is in the +2-oxidation state and stereochemical active Ge 4$s$ lone pair is present and oriented opposite to the triangle of $Se^{2-}$ anions, Figure 2 [30]. The energy threshold $E_0$ position is taken at half-height of the normalized absorption edge. According to dipole selection rules in the one-electron consideration, the Ge K-absorption edge (sharp rise in XANES spectrum + A feature) is due to the allowed transition from the Ge 1$s$ core state to the first empty density of states (DOS) involving the $p$(Ge) atomic orbitals in the solid state [31]. The electronic density of the unfilled states is governed by the number and the nature of the chemical bonds between germanium and its first neighbors. In α-GeSe, the Ge atom "sees" three Se atoms as first neighbors forming Ge-Se heteropolar bonds (~2.58 Å) with a weakly ionic character. The broad B feature and the C shoulder, located around 7-9 eV above the white line A, are more likely associated with a peak in the solid-state density, primarily influenced by the surrounding Ge geometry (multiple-scattering resonance), rather than the electronic properties.

The XANES features broadening and the A intensity reduction in zones I and II compared to zone III, Figure 4(b), are due to the typical structural disorder effect in the amorphous state.

For each experimental loop, after having registered the XRS pattern, then the Ge K-edge XANES spectrum, the fluorescence-detected XANES spectrum was collected at the Se K-edge, Figure 5, with continuous heating, which explains the slight temperature variations between XRS and XANES data (Figures 1, 4 and 5).

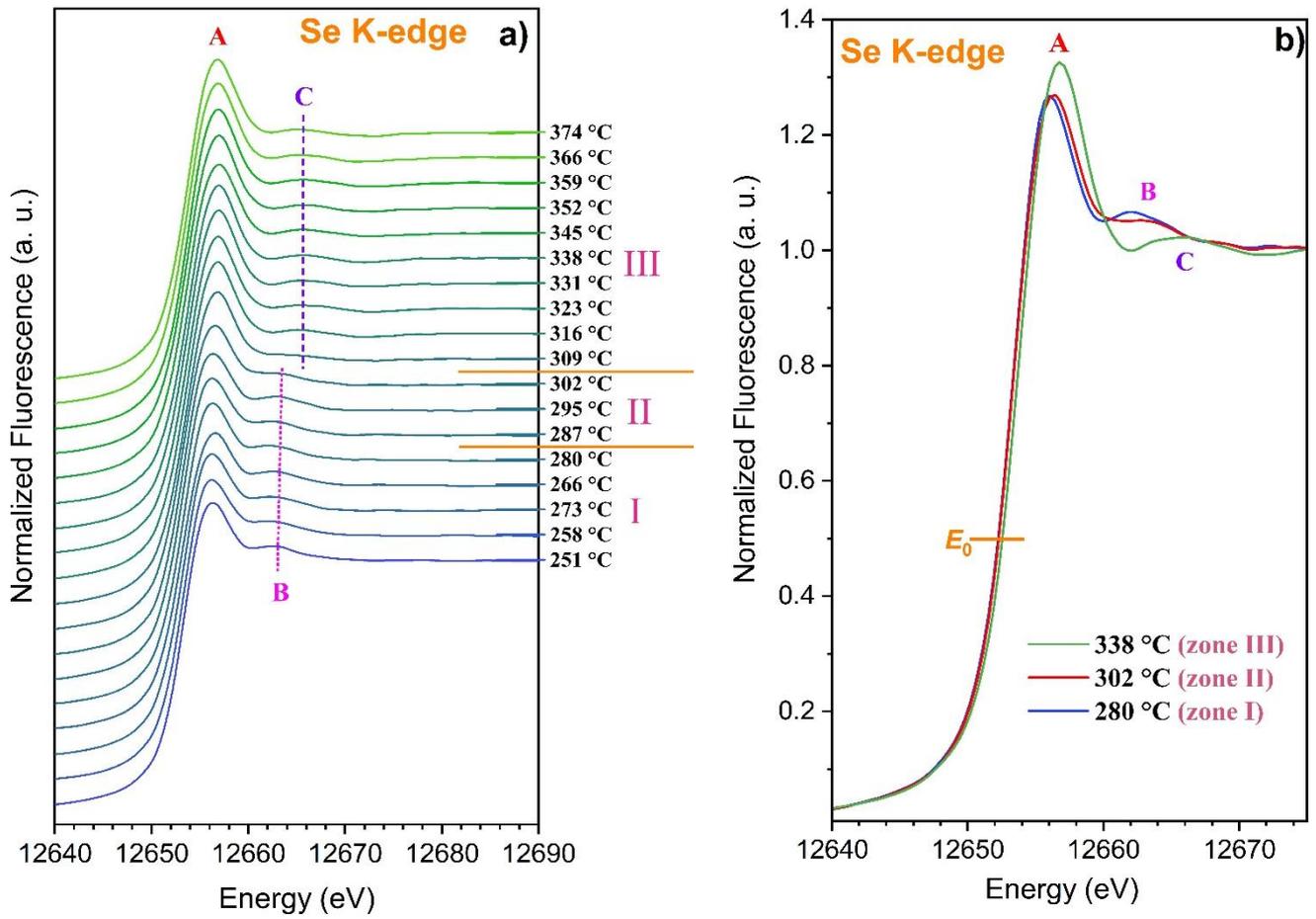

**Figure 5:** a) Evolution with the heating temperature registered in isochronal annealing conditions of the near-edge features in the fluorescence spectra registered at the Se K-edge of the as-deposited amorphous GeSe thick film. The spectra have been stacked for clarity. b) XANES spectra at selected temperature. The heating rate was 0.2 °C/min.

The ground state electronic configuration of the Se atom is $[Ar]4s^2 3d^{10} 4p^4$. Se K-edge XANES probe the final unoccupied partial density of states with dominant $p$-like symmetry. In the amorphous state, zone I in Figure 5(a), the white line A is located at 12656.1 ±0.5 eV, and the broad feature B, multiple scattering of the photoelectron with the near neighbors, at 12662.2 ±0.5 eV. The energy of the $1s$ core level is $E_0$ = 12653.0 ±0.5 eV. In the crystalline state of zone III with the α-GeSe structure, we always have a white line (WL) feature located at 12656.7 ±0.5 eV followed by a multiple scattering feature called C at 12665.6 ±0.5 eV. In this zone, $E_0$ = 12653.7 ±0.5 eV. In α-GeSe, Se atoms are four-fold coordinated with Ge, forming 2 short Se-Ge bonds and two long Se—Ge nonbonding interactions, Figure 2. For zone II, where there is a coexistence of the amorphous and the crystalline state, the general aspect of the XANES spectra from T = 287 °C to T = 302 °C presents a close resemblance to those in zone I, the amorphous part is dominant. For T = 309 °C, the crystalline contribution is dominant, and thus, the Se K-edge XANES spectrum looks like those in zone III where only the α-GeSe structure exists with Se having a nominal oxidation state of −2.

Figure 5(b), illustrates i) the increase in amplitude of the white line A even with the thermal damping of the Debye-Waller coefficient, and ii) its higher-side energy displacement with the amorphous to the crystalline temperature-induced transition of the GeSe sample. The B feature's disappearance and the C one's formation from amorphous to crystalline states indicate different multi-scattering pathways.

### 3.2 GeSe$_{0.75}$Te$_{0.25}$ substituted composition

#### 3.2.1 Temperature-dependent X-ray scattering patterns

*In-situ* temperature-dependent synchrotron XRS patterns registered in isochronal annealing conditions on a GeSe$_{0.75}$Te$_{0.25}$ 3-μm-thick film are presented in Figure 6. A first onset of crystallization appears at $T_{c1}$ = 242 °C and a second at $T_{c2}$ = 257 °C, Figure 6(a). The XRS patterns can be divided into 5 distinct zones depending on the heating temperature. In zone I, only diffusion rings are visible which are the signature of the amorphous state of the film. Zone II, from $T_{c1}$ to 250 °C, is a mixing of an amorphous and a crystalline phase C1 whose proportions evolute with T. Above 250 °C, only diffracted peaks of phase C1 are present in zone III up to the apparition of new diffracted peaks at $T_{c2}$ corresponding to phase C2. In zone

IV [257 °C to 269 °C], the two phases C1 and C2 coexist while only phase C2 is present in zone V with T > 269 °C, Figures 6(a) and 6(b).

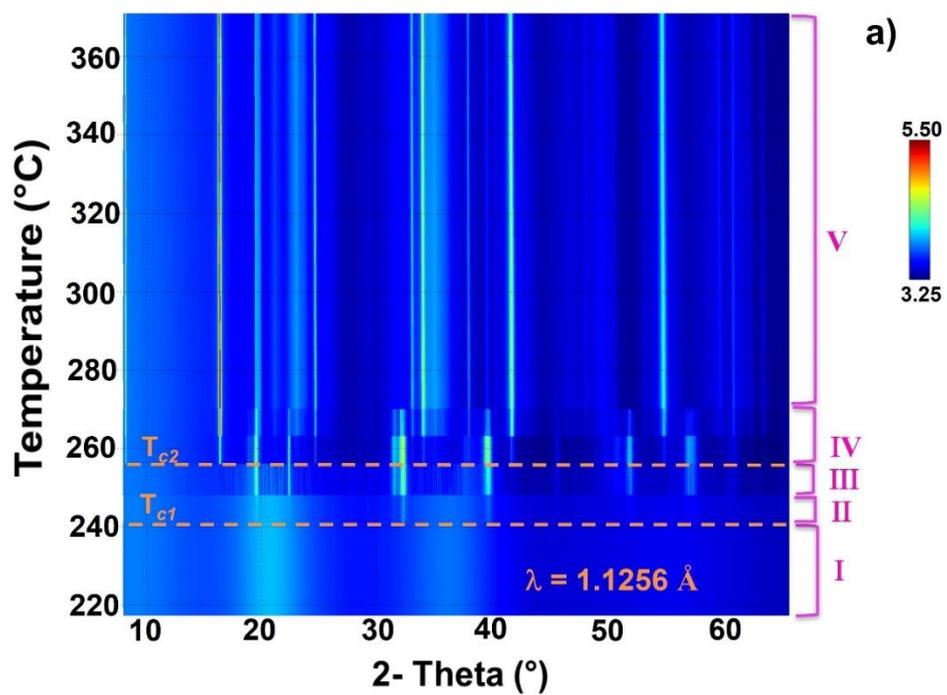

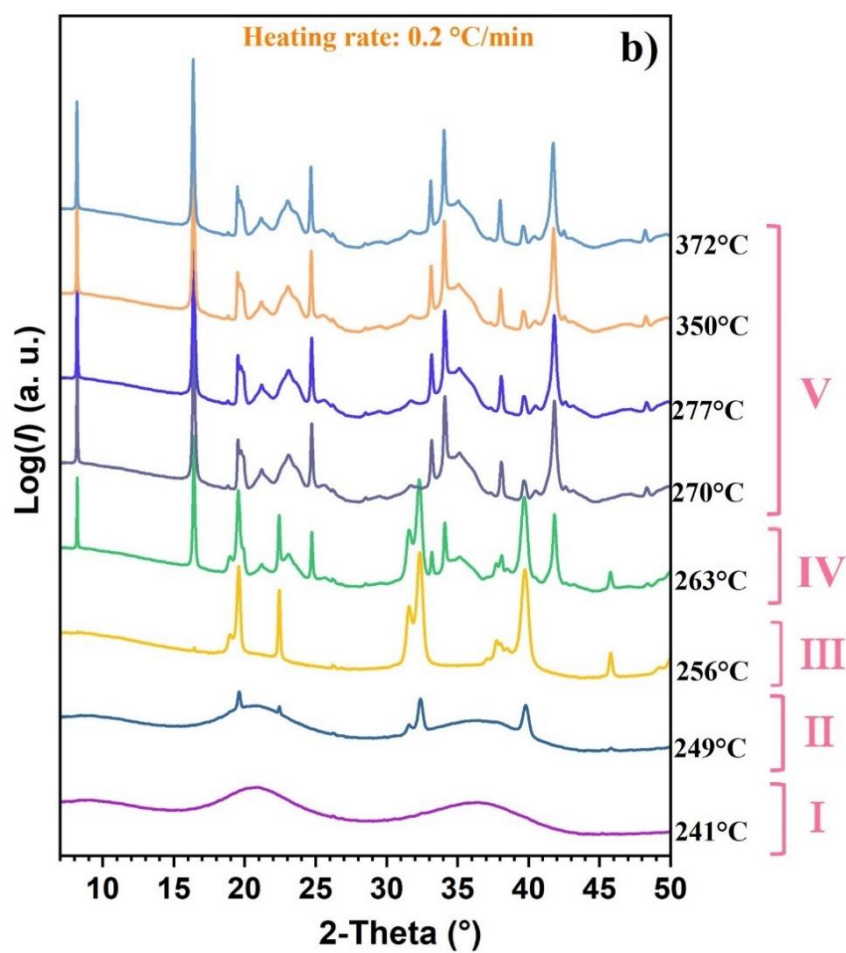

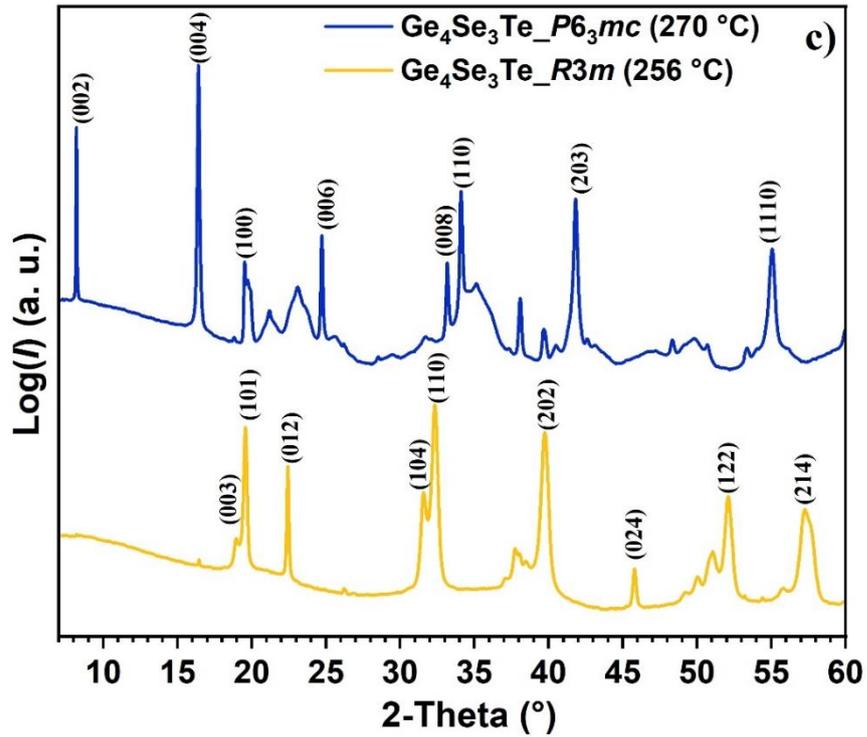

**Figure 6:** a) *In-situ* temperature-dependent synchrotron XRS patterns registered in isochronal annealing conditions on a GeSe$_{0.75}$Te$_{0.25}$ 3-μm-thick film deposited on a Si(001) substrate (log scale in color). b) XRS patterns at selected temperatures. The spectra have been stacked for clarity. c) XRD patterns of the GeSe$_{0.75}$Te$_{0.25}$ crystalline phase with a rhombohedral *R*3*m* and a *P*6$_3$*mc* hexagonal structure. Only the main reflections were indexed on the graph. The heating rate was 0.2 °C/min.

The heating of the Te 25 at. % doped GeSe 3-μm amorphous thick film, with our experimental conditions, leads to the crystallization of phase C1 that corresponds to the metastable rhombohedral structure of Ge$_4$Se$_3$Te (GeSe$_{0.75}$Te$_{0.25}$) with the low symmetry *R*3*m* [19, 32]. This phase, isostructural of the well-known phase change material α-GeTe [4, 5], is present as a single crystallized phase only over a narrow temperature range, Figure 6(a). The diffraction pattern registered with λ = 1.1256 Å is shown in Figure 6(c) for T = 256 °C. It is composed of distorted octahedral coordination of Ge by *X* = Se/Te, Figure 7, with 3 short heteropolar bonds (*d*~2.72 Å, green line) and 3 longer nonbonding interactions (*d*~3.05 Å, bicolor cylinders) forming a [3+3]-coordination (distorted octahedral Ge*X*$_6$ site) as in α-GeTe.

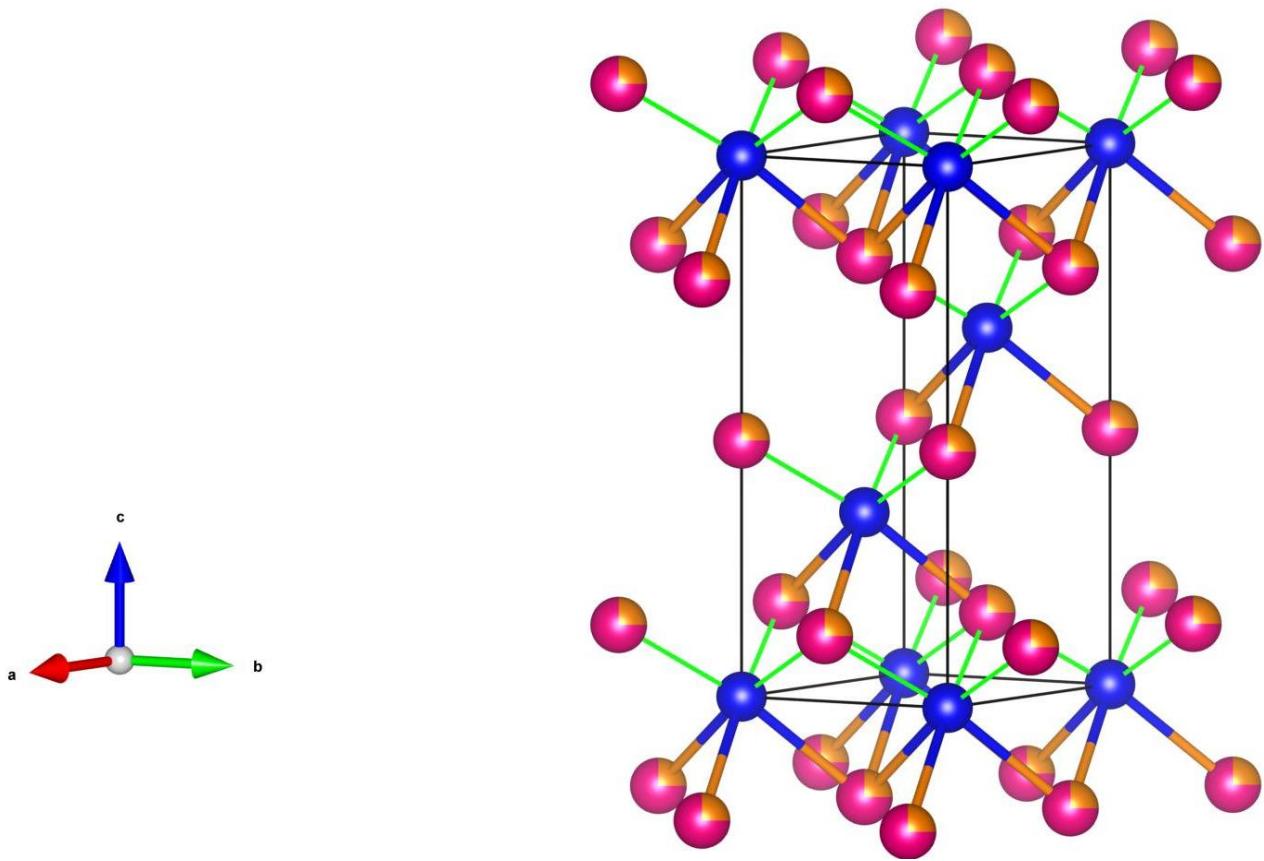

**Figure 7:** Crystal structure of the rhombohedral polymorph of Ge$_4$Se$_3$Te (space group $R3m$) representing in the hexagonal setting. The black line is the unit cell. Ge in blue, Se in pink, and Te in gold balls. The 3 first-nearest Ge-Se/Te bonds are shown with green lines, and the 3 next-nearest nonbonding lengths (bicolor cylinders) are also shown. Only the first coordination shell of Ge atoms is depicted in the figure for clarity but an identical distorted octahedral environment exists for the Se/Te atoms.

From T$_{C2}$ = 257 °C, we notice the apparition of diffraction peaks in addition to those already present, due to forming a second phase C2 with a different structure. This second phase is the stable hexagonal polymorph of the Ge$_4$Se$_3$Te composition with the $P6_3mc$ structure [18, 19, 34]. Its diffraction pattern shown in Figure 6(c) was registered at 270 °C. h-Ge$_4$Se$_3$Te is formed by 2D layers *A*-Ge-Ge-*A* (*A* = 3:1 ratio of Se and Te) along the *c*-axis separated by van der Waals-like gaps, Figure 8 [18, 19]. Ge is coordinated in a distorted octahedron of *A* (3\*Ge-*A* at ~2.60 Å) and Ge atoms (3\*Ge-Ge at ~2.92 Å). Within one *A*-Ge-Ge-*A* layer, the covalent type coupling is strong and short Ge...Ge contacts stabilize the crystal structure.

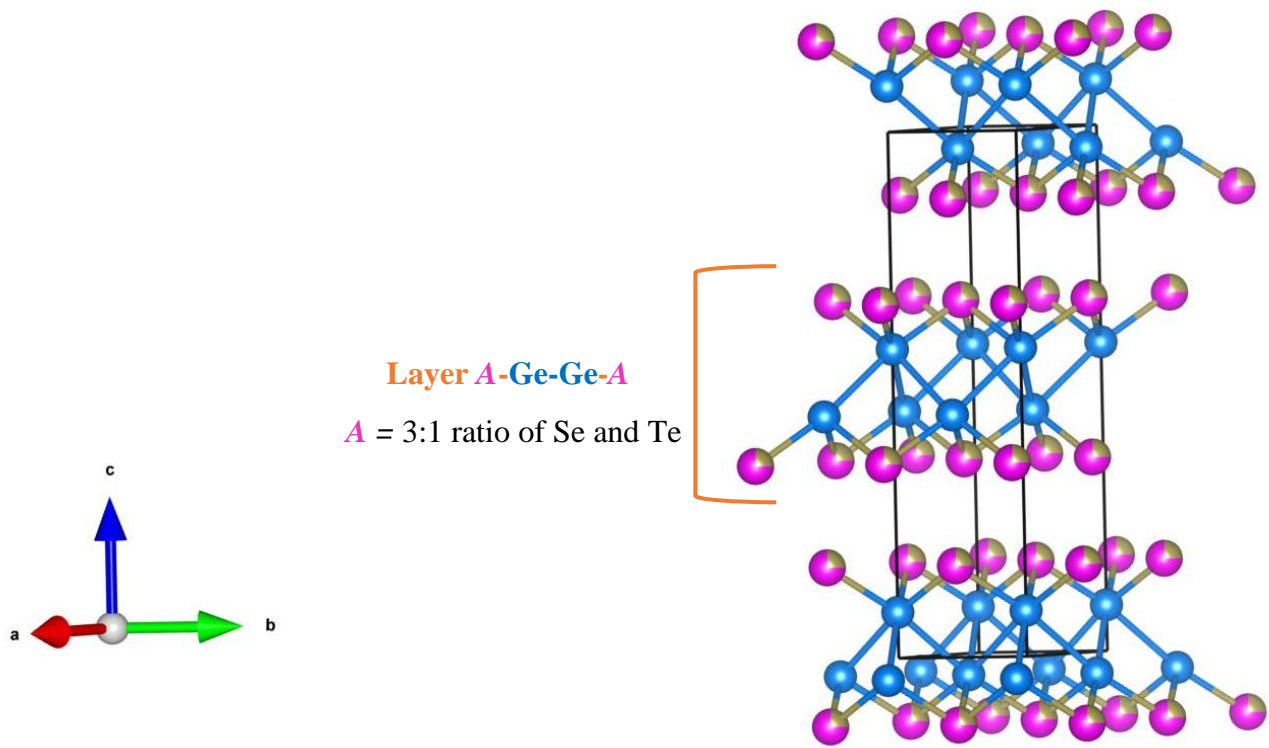

**Figure 8:** Crystal structure of the hexagonal polymorph of Ge$_4$Se$_3$Te (space group *P6$_3$mc*). The black line is the unit cell. Ge in blue, Se in pink, and Te in gold balls. The 3 first-nearest Ge-(Se/Te) bonds are shown with bicolor cylinders and the Ge—Ge contacts are in blue. Only the first coordination shell of Ge atoms is depicted in the figure for clarity.

*3.2.2 Temperature-dependent XANES spectra*

Figure 9(a) shows the results of the measurements of the Ge K-absorption edge, on heating, of the Te doping GeSe thick film. Ge-K XANES spectroscopy directly probes the empty 4*p*-states of Ge atoms. The absorption curves exhibit significant variations in temperatures.

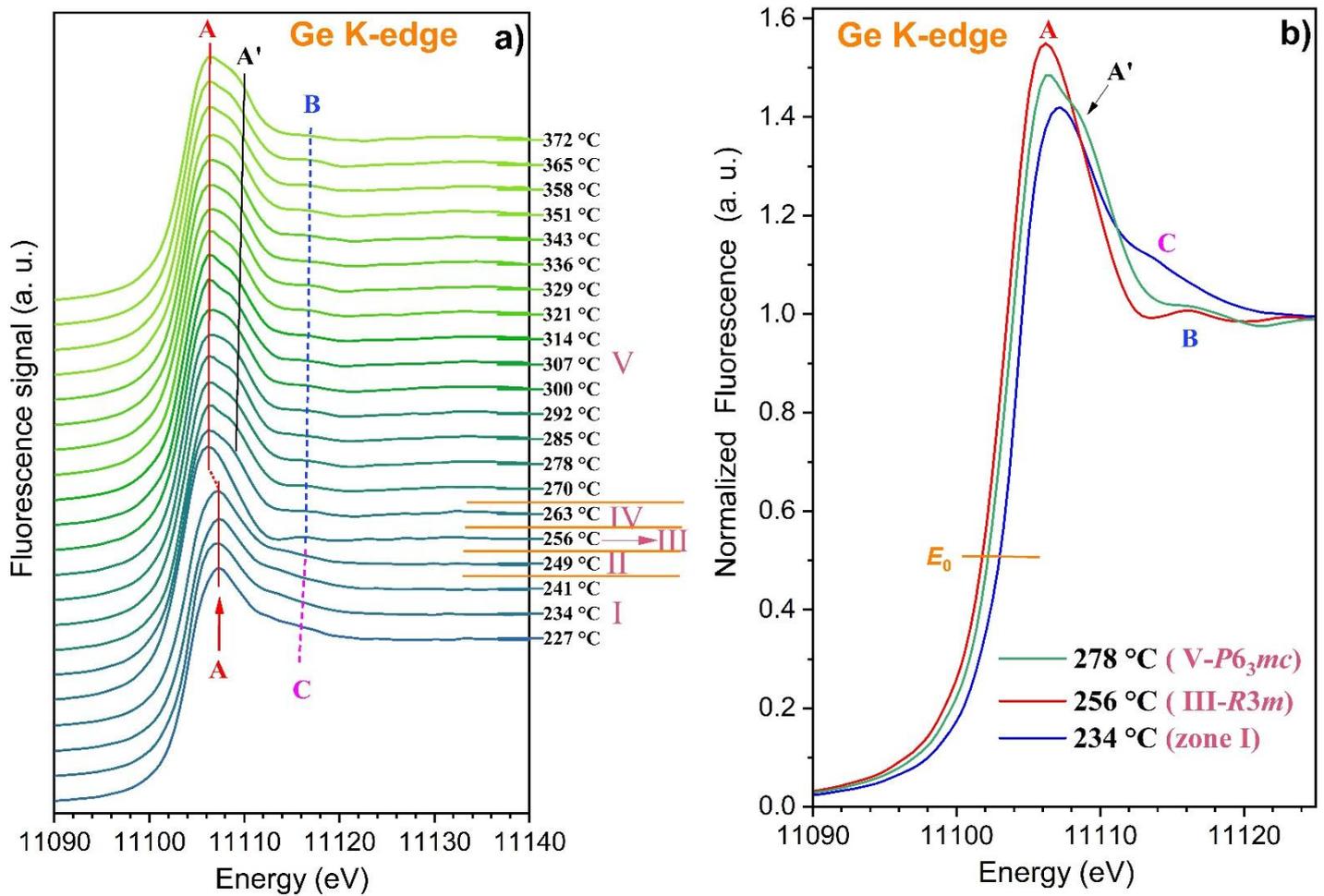

**Figure 9:** a) Evolution with the heating temperature registered in isochronal annealing conditions of the normalized near-edge features in the fluorescence spectra registered at the Ge K-edge of the as-deposited amorphous $GeSe_{0.75}Te_{0.25}$ thick film. The spectra have been stacked for clarity. b) Normalized spectra at selected temperatures. The heating rate was 0.2 °C/min.

The Ge K-edge absorption spectra in zone I, corresponding to the amorphous state, present a WL peak (indicated by A) with a maximum at 11107.3 ±0.5 eV, and a shoulder labeled C at 11113.7 ±0.5 eV, Figure 9(b). The same general aspect of the recorded fluorescence spectrum is found in zone II corresponding to the mixing of the amorphous state with the onset of crystallization. The XANES spectrum collected at T = 256 °C, zone III, is the fingerprint of the metastable crystalline $Ge_4Se_3Te$ phase with the rhombohedral structure isotype of α-GeTe ($R3m$). We notice a clear evolution of the XANES shape of zone III compared to that in zone

I: the A resonant feature is shifted towards lower energy (11106.2 ±0.5 eV) and its intensity is higher, the C shoulder is no longer present, and a bump, labeled B, appears at 11116.0 ±0.5 eV. For this r-Ge$_4$Se$_3$Te compound, the position of $E_0$ (11102.0 ±0.5 eV) is shifted to the lower side of the energy scale compared to zone I.

Still raising the temperature, the stable hexagonal polymorph of Ge$_4$Se$_3$Te (*P6$_3$mc*) starts to be formed, zone IV, whose proportion increases as the content of the r-Ge$_4$Se$_3$Te phase decreases. From $T$ = 270 °C, only the hexagonal polymorph is present up to the maximum recorded temperature. Its Ge K-edge XANES spectrum in zone V consists of a broad post-edge feature (white line) with two maxima, A (at 11107.0 ±0.5 eV for T = 278°C, Figure 9(b)) and A' (11108.6 ±0.5 eV), and a broadened B bump centered at 11116.8 ±0.5 eV. Its 1*s* core excitation level $E_0$ equals 11102.6 ±0.5 eV at T = 278 °C. The Ge first coordination shell in h-Ge$_4$Se$_3$Te material contains 3 short Ge-*X* bonds (*X* = Se/Te) at ~2.60 Å and 3xGe...Ge short interactions at ~2.92 Å, see Figure 8. The first maxima A is thus attributed to the Ge-*X* bonding and the second one A' to Ge...Ge short interactions, Figure 9(a). On the opposite, the white line of the ternary amorphous sample at the Ge edge is thinner and presents only one maximum A.

Se K-edge XANES spectra registered for the Te-doped GeSe parent film are compared for selected temperatures in Figure 10. The small temperature variations between XRS and XANES data are due to the continuous heating condition selected for these synchrotron experiments.

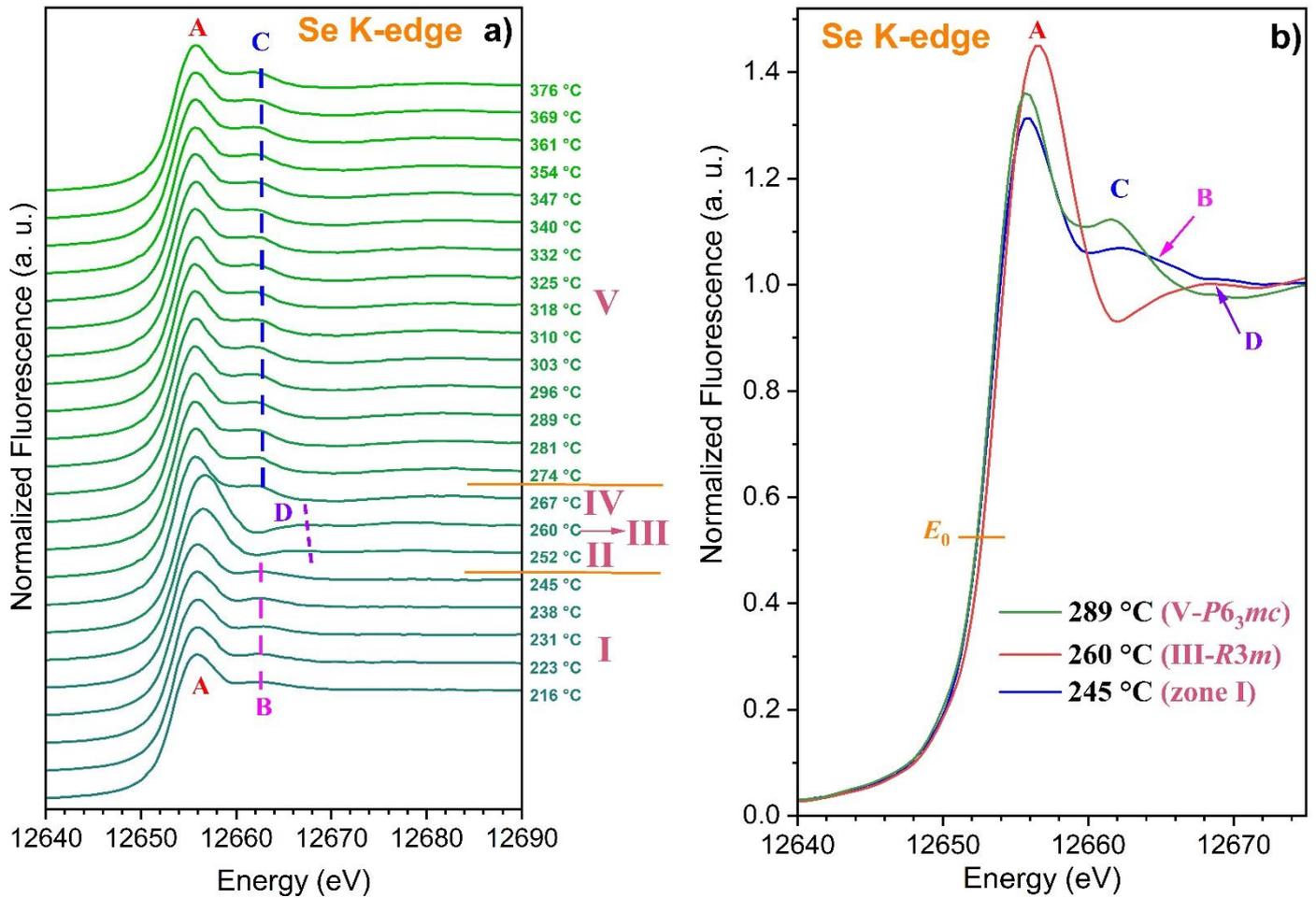

**Figure 10:** a) Evolution with the heating temperature registered in isochronal annealing conditions of the fluorescence spectra registered at the Se K-edge of the as-deposited amorphous $GeSe_{0.75}Te_{0.25}$ thick film. The spectra have been stacked for clarity. b) XANES spectra at selected temperatures. The heating rate was 0.2 °C/min.

Three distinctive experimental spectral shapes are visible in Figure 10(a). The one in zone I corresponds to the amorphous state of $GeSe_{0.75}Te_{0.25}$ with an A white line and a large bump B, Figure 10(b). The second shape with a white line A and a D broad hump, found in zones II and III, is the fingerprint of r-$Ge_4Se_3Te$ (*R3m*). Then, the third general XANES shape found in zones IV and V is the fingerprint of the stable hexagonal polymorph of $Ge_4Se_3Te$, Figure 10(b). The values of the identified features in Figure 10(b) and the energy position of the 1*s* core level are gathered in Table II. As for the Ge K-edge, the fluorescence-type XANES spectra recorded at the Se K-edge present different general shapes from the amorphous state to the crystallized ones indicating modified electronic and geometric structure.

**Table II:** Energy position in eV of the 1s core level $E_0$ and the main XANES features registered at the Se K-edge of the GeSe$_{0.75}$Te$_{0.25}$ sample in function of the heating temperature.

|  | **T = 245 °C** | **T = 260 °C** | **T = 274 °C** |
|---|---|---|---|
|  | *zone I* | *zone III* | *zone V* |
|  | amorphous state | crystalline state Ge$_4$Se$_3$Te (*R3m*) | crystalline state Ge$_4$Se$_3$Te (*P6$_3$mc*) |
| **A** | 12656.0 ±0.5 | 12656.8 ±0.5 | 12655.7 ±0.5 |
| **B** | 12662.3 ±0.5 |  |  |
| **C** |  |  | 12662.3 ±0.5 |
| **D** |  | 12667.6 ±0.5 |  |
| ***E$_0$*** | 12652.2 ±0.5 | 12653.0 ±0.5 | 12652.6 ±0.5 |

## 4. Discussion

Germanium selenides stand as the archetypal chalcogenide glass-forming system, subject to thorough scrutiny of their structural characteristics over the past twenty years through various experimental and computational studies. However, to date, no univocal determination of the structural properties of amorphous GeSe, at ambient conditions, is available [35 – 40]. The diversity of the local order and structural arrangement of amorphous GeSe exhibit a range of structures that may involve various configurations of Ge and Se atoms. Two main coordination number models of the Se and Ge atoms in amorphous GeSe stoichiometric composition are found in the literature: i) Ge and Se atoms predominantly four- and two-fold coordinated (4(Ge):2(Se)), respectively, with each atom satisfying its normal valence requirement [38, 40]. With this local arrangement, homopolar Ge-Ge pairs would be necessary in addition to heteropolar Ge–Se pair to maintain the charge neutrality, and ii) a 3-fold coordination model (3(Ge):3(Se)) as in the corresponding α-GeSe crystal with only heteropolar Ge-Se bonds, see Figure 2 [36, 37].

Under our experimental conditions, the onset of crystallization of the thick film with the GeSe parent composition occurred at $T_c$ = 284 °C and completely crystallized at T > 305 °C, Figures 1(a) and 1(b). The high-temperature X-ray diffraction patterns in the 312-370 °C range confirm the phase purity with only diffraction peaks associated with the layered α-GeSe material. Moreover, in this temperature range, no significant changes in the intensity ratios of the reflections and no discontinuity of the thermal expansion of both lattice parameters and unit-cell volumes, Figure 3, are observed, in agreement with the lack of a phase transition. Thermal expansion of α-GeSe has an anisotropic character, Figure 3(a). The slight linear thermal expansion along the *a* and *b* orthorhombic axes in the 312-370 °C range agrees with previous works [28, 41]. A zero thermal expansion along *c* is reported in [41], supporting the present measurements, while a negative expansion is shown in [28].

When replacing 25 at.% of Se atoms with Te ones in the amorphous GeSe parent composition, the onset of crystallization decreased to $T_{c1}$ = 242 °C with our experimental conditions. The decrease of $T_c$ with the Se/Te substitution indicates the diminution of the thermal stability of the unstable amorphous state. Indeed, the partial replacement of small-size Se (119 pm) by giant Te (142 pm) leads to a local surrounding distortion that decreases the rigidity of its structural network compared to the undoped GeSe compound. Thus, less thermal energy is necessary to activate long-term crystallization from the glassy state.

According to the phase equilibria in the GeTe-GeSe pseudo-binary system established from bulk samples [21, 22, 34], the GeSe$_{1-x}$Te$_x$ (0.15 ≤ x ≤ 0.39) alloys crystallize in a single-phase of general composition GeSe$_{0.75}$Te$_{0.25}$ (Ge$_4$Se$_3$Te) with a hexagonal structure (*P6$_3$mc* space group) stable only below ~ 400 °C [18, 19, 33, 34]. However, when continuously heated under our experimental conditions, from room temperature to $T_{max}$ = 372 °C, the GeSe$_{0.75}$Te$_{0.25}$ thick film deposited on Si substrate switched between the amorphous and crystalline phases in a two steps process: first, the amorphous phase transformed at $T_{c1}$ to an intermediate metastable crystallized phase of Ge$_4$Se$_3$Te composition with a noncentrosymmetric *R3m* space group. The crystallization of the metastable r-Ge$_4$Se$_3$Te alloy was already shown after heating an amorphous GeSe$_{0.75}$Te$_{0.25}$ thick film deposited on a glass substrate (this work = Si substrate) with a heating rate of 2 °C/min and 0.2 °C/min (this work = 0.2 °C/min) [33]. This rhombohedral structure exists only because the as-deposited thick film is out of equilibrium during the thermal treatment. Then, a second phase transition started at $T_{c2}$ ($T_{c2}$ > $T_{c1}$) from the polar rhombohedral-type to the layered hexagonal-type structure of Ge$_4$Se$_3$Te with space group *P6$_3$mc*.

Comparing our *in-situ* transition temperatures to those in [33] for a GeSe$_{0.75}$Te$_{0.25}$ 3 μm thick film deposited *via* the same experimental set-up, we noticed that $T_{c1}$ and $T_{c2}$ depend on the temperature ramp during the synchrotron XRS data recording and the substrate's type. When the ternary alloy sample is deposited on a Si substrate and heated at the rate of 0.2 °C/min (this work), the r-Ge$_4$Se$_3$Te polymorph is detected only in a small temperature range, $T_{c1}$=242 °C and $T_{c2}$ = 257 °C ($\Delta T_c$ = 15 °C), Figure 6(a). When deposited on a glass substrate [33] and heated with a rate of 2 °C/min, $T_{c1}$= 270 ± 2 °C and $T_{c2}$ = 318 ± 2 °C ($\Delta T_c$ = 48 °C) while $T_{c1}$= = 249 ± 2 °C and $T_{c2}$ = 263 ± 2 °C ($\Delta T_c$ = 14 °C) at a rate of 0.2 °C/min. With the slower heating rate, the structural evolution tends towards a quasi-static transformation and the system has time to reorganize, thus the domain of existence of the metastable rhombohedral phase r-Ge$_4$Se$_3$Te decreases. Considering the two substrates, a global shift of the crystallization temperatures is observed when the sample is heated at the same rate, but the temperature range of the existence of the r-Ge$_4$Se$_3$Te polymorph is the same (14 °C). These temperature shifts observed in function of the nature of the substrate used are certainly related to their different thermal properties as the thermocouple during the *in situ* XRS experiments, this work and [33], was encapsulated in a heating plate localized under the "substrate-film" sample.

No indication of crystallization of individual elements (segregation) or parasitic phases was found in the crystallized state whatever the recorded temperature. XANES is a highly effective experimental technique for distinguishing between Ge–$X$ ($X$ = Te, Se) and Ge–O bonds in GeSe-based compounds, thanks to the significant shift of approximately 7 eV in the Ge K-edge X-ray absorption energy between Ge–X (Ge$^{+II}$) and Ge–O (Ge$^{+IV}$) bonds [42]. Our temperature-dependent Ge K-edge X-ray fluorescence experiments, conducted under an inert atmosphere, revealed no evidence of Ge-O bond contribution in our uncapped thick film samples, irrespective of whether they were in the amorphous or crystalline state. The XANES data confirms the lack (within the detection limit) of oxidation on the surface of our samples i.e., the formation of a GeO$_2$ layer.

Temperature-induced variations observed in Ge and Se K-edge XANES spectra of GeSe parent, Figures 4 and 5, indicate adjustments in the network's short and intermediate-range structural ordering. Figure 4(b) reveals the temperature dependence of the energy variation in the Ge K-edge position of GeSe film from amorphous to crystalline states with a gradual shift of the rising absorption edge ($E_0$ position of half height (0.5) of normalized XANES spectrum) towards lower energies of about −0.8 ±0.5 eV. This small temperature-dependent displacement does not suggest a change in the oxidation state of Ge. Still, it is the signature of a gap closure process under temperature. The Fermi level in the amorphous GeSe sample would be higher in energy than in the anisotropic layered semiconductor α-GeSe whose band gap is announced at an energy of 1.21 ± 0.05 eV at room temperature [43].

Furthermore, the α-GeSe crystallization is accompanied by an amplitude increase of the near-edge structure A (even with the thermal damping), due to a greater Ge 4*p*-type electron unfilling, and by a broad constructive interference B apparition in the post-edge region. B is a multiple scattering feature, as more than one backscatter can contribute to the XANES signal. As edge, near-edge and post-edge XANES are susceptible to the geometric details of the absorbing site and its short- and medium-range orders, the registered XANES evolutions from amorphous to the crystalline state of the GeSe film, reflect modifications of the Ge surrounding. At the Se K-edge, variations of the Se surrounding between the amorphous and the crystalline state of GeSe are also found, Figure 5(b).

For samples in the amorphous state, when replacing 25 at. % of Se atoms by Te ones in the GeSe parent composition, the general aspect of the Ge K-edge fluorescence signal is close to that of undoped GeSe, Figure 11(a). The broad features are consistent with the disordered structure of the samples. The major differences are in the $E_0$ and A energy position and the A amplitude, Table III.

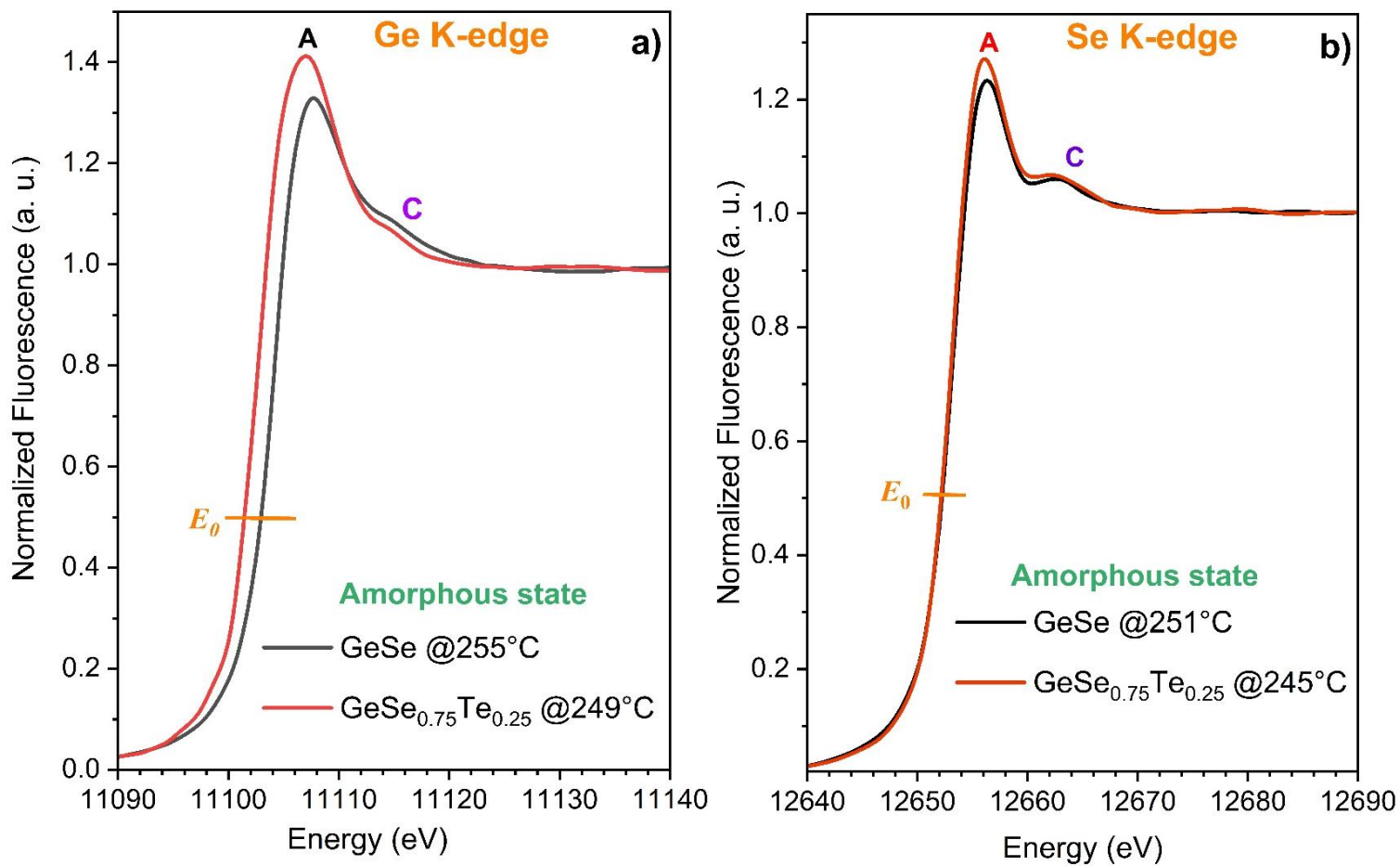

**Figure 11:** Evolution with the chemical composition of the near-edge features in the fluorescence spectra registered at the a) Ge K-edge and b) Se K-edge.

**Table III:** Energy position in eV of the 1*s* core level $E_0$ and the main XANES features registered at the Ge, Se K-edges of the GeSe and GeSe$_{0.75}$Te$_{0.25}$ samples in the amorphous state.

|  | $E_0$ | A white line | C shoulder |
|---|---|---|---|
| **Ge K-edge** | | | |
| **GeSe** | 11103.6 ±0.5 | 11107.7 ±0.5 | 11115±0.5 |
| **GeSe$_{0.75}$Te$_{0.25}$** | 11102.4 ±0.5 | 11107.0 ±0.5 | 11115±0.5 |
| **Se K-edge** | | | |
| **GeSe** | 12652.4 ±0.5 | 12656.1 ±0.5 | 12662.8 ±0.5 |
| **GeSe$_{0.75}$Te$_{0.25}$** | 12652.2 ±0.5 | 12656.0 ±0.5 | 12662.3 ±0.5 |

When comparing XANES data registered at close temperature, the Ge K-edge threshold energy position $E_0$ for the amorphous Te-doped phase is shifted by approximately 1.2 ±0.5 eV to the lower energy side compared to that of the amorphous GeSe parent composition, Figure 11(a) and Table III. This energy shift of a core state is correlated to an energy decrease of the conduction band and thus, to a smaller band gap of the GeSe$_{0.75}$Te$_{0.25}$ phase. For an amorphous 500-nm thick GeSe film, the optical band gap $E_g$ value is 1.37 eV at 50°C and ~1 eV for the doped composition [32]. A decrease in calculated Fermi energy $E_F$ from GeSe to GeTe compositions was also reported in [38]. The Te doping is responsible for the Ge K-edge threshold shift due to the distribution of the atom's overall charge *via* the presence of a different degree of bond covalency, and/or number of bonds, and/or distances from the central atom which influence the bonding *p*-like electron originating from the Ge atom in the density of state (DOS). Indeed, a XANES spectrum shows a lower energy shift when the absorber atom carries a less positive charge, decreasing the Coulomb interaction between the nuclear charge and core electrons. Thus, less energy is required to excite an electron from an inner orbital. A lone pair as the Ge 4*s* electron pair can influence the charge distribution by either repelling or attracting electrons.

A broadening of the A feature is also seen for the ternary alloy, Figure 11(a), linked to a structural disorder augmentation. The random replacement of 25 % of Ge-Se bonding distances by Ge-Te bonds induces a modification in the local environment of the Ge atoms (bond lengths, angles spread,…) and thus, less localized *p*-type empty states.

A resemblance in the general aspect of the Se K-edge XANES curves is seen in Figure 11(b) between GeSe and Te-doped GeSe amorphous compositions indicating close Se-type partial DOS of *p* orbital contributions and local geometry around the Se atoms. The energy positions of the Se threshold, the white line A, and the C multi-scattering features are given in Table III. From the high similitude in the general aspect of the XANES spectra at the Ge and Se K-edges of the two amorphous compositions under study, a close Ge and Se geometrical surroundings could be considered.

Heating the amorphous $Ge_4Se_3Te$ ($GeSe_{0.75}Te_{0.25}$) thick film, the first crystallized phase to appear is the metastable rhombohedral r-$Ge_4Se_3Te$ one, Figure 6, isotype of α-GeTe, (Ge and Se [3+3]-coordination shell, Figure 7). The Ge and Se K-edge XANES spectra of these two phases, registered at different temperatures and compared in Figures 12(a) and 12(b) respectively, present strong feature differences. This is an indication of electronic and local structure evolutions from the amorphous to the layered crystallized phase. The edge and post-edge spectral signatures are related to the nearest and next-nearest neighbor modifications associated with the absorbing atoms' geometry. The sharp white line A amplitude, for the two K-edges, is higher in the crystalline state even with the increase of the registered temperature and consequently of the thermal dumping (Debye-Waller coefficient). The amplitude is influenced by several factors linked to the electronic structure and thus, to the chemical environment of the absorbing atoms. We may argue that the first-shell coordination number in the amorphous GeSe$_{0.75}$Te$_{0.25}$ sample is less than in the layered rhombohedral phase which is 6, Figure 7.

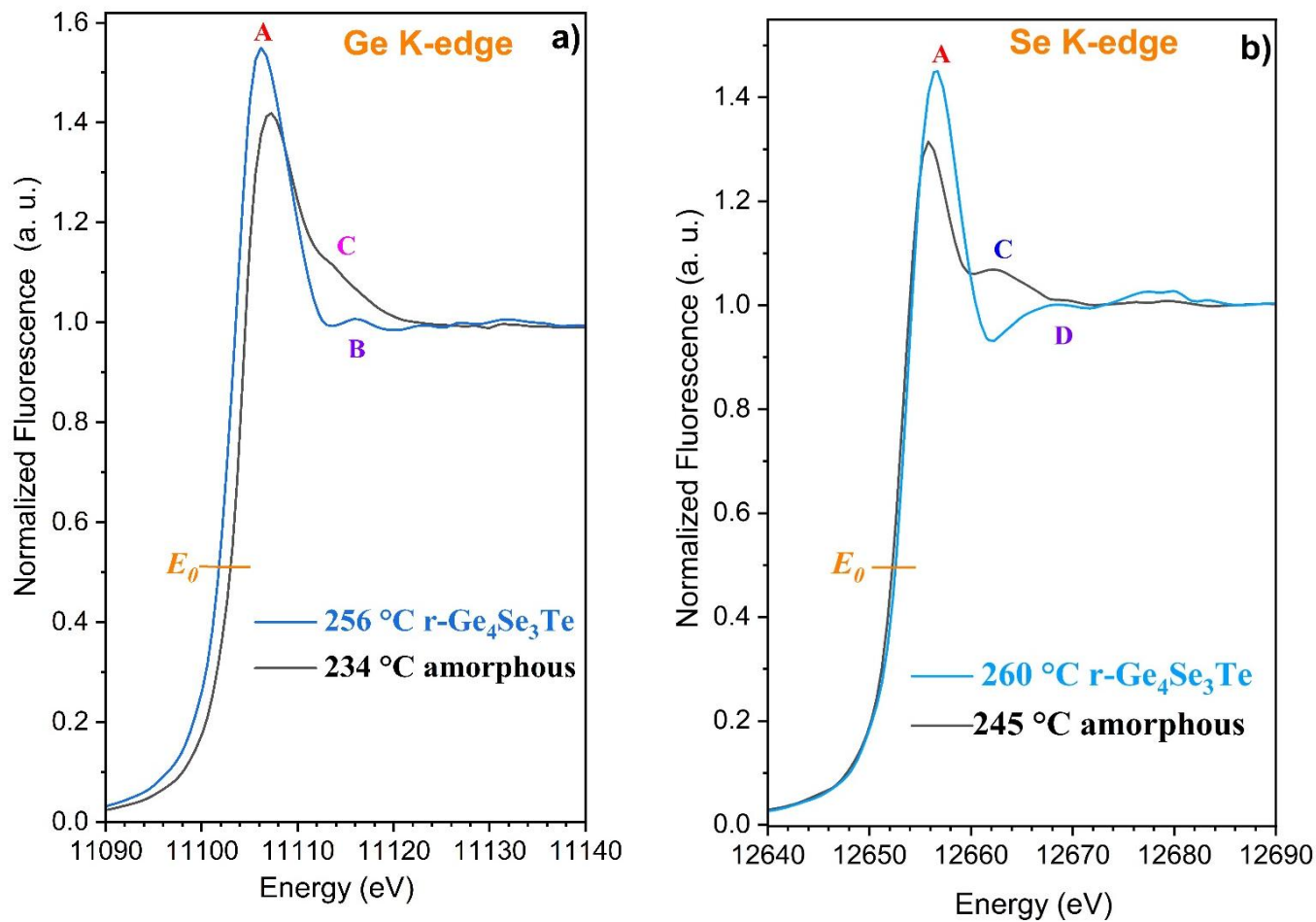

**Figure 12:** Evolution with the nature of the state (amorphous or *R*3*m* crystallized) of the near-edge features in the fluorescence spectra registered for the nominal GeSe$_{0.75}$Te$_{0.25}$ composition at the a) Ge K-edge and b) Se K-edge.

## 5. Conclusions

**In situ, temperature-dependent structural insights into GeSe and GeSe$_{0.75}$Te$_{0.25}$ compositions, deposited as 3 µm-thick amorphous films via the co-evaporation technique, were obtained using two complementary characterization methods: X-ray scattering, which probes long-range structural order and disorder, and XANES, which provides sensitivity to both the unoccupied electronic density of states and the local atomic order/disorder around the targeted elements.**

For the amorphous GeSe parent composition, crystallization began at $T_c$ = 284 °C and was completed at 305 °C, with the formation of the α-GeSe structure (*Pnma* space group), having a positive linear thermal expansion along the orthorhombic *a*- and *b*- axes and a zero thermal expansion along the *c*-axis. The layered α-GeSe crystallizes in a preferential crystallographic direction belonging to the (*h*00) planes (*h* pair) and is stable up to 370 °C, the maximum temperature registered in this work. Substituting 25 at.% of Se by Te reduced the crystallization onset to $T_{c1}$ = 242 °C, indicating a diminution of the amorphous thermal stability due to bonding and structural changes induced by the presence of giant Te atoms. Under *in situ* continuous heating, the amorphous GeSe$_{0.75}$Te$_{0.25}$ thick film first transitioned at $T_{c2}$ = 257 °C to a metastable rhombohedral compound (r-Ge$_4$Se$_3$Te, *R3m*) isostructural of α-GeTe, the well-known PCM alloy, followed by the formation of the stable hexagonal polymorph (h-Ge$_4$Se$_3$Te, *P6$_3$mc*) at higher temperatures. Furthermore, XRS data supported the absence of crystalline parasitic phases or Ge/Se/Te clusters for the two compositions during heating.

The structural disorder of the amorphous state is reflected on the XANES spectra by the broader and reduced intensity of the resonances compared to the crystalline state. Moreover, Ge K-edge white line is larger in the Te-substituted amorphous film compared to the amorphous GeSe, due to the co-existence of at least Ge-Te and Ge-Se heteropolar bonds that increase the average distances and angle spread. Mixing of Se and Te atoms in the ternary alloy led to an energy decrease of the Ge K-edge threshold while no shift was shown for the Se K-edge and an increase of the white line amplitude. This reflects a variation in the Ge empty DOS compared to that of GeSe that is correlated to the modification of the average degree of covalency and Ge local surroundings in $GeSe_{0.75}Te_{0.25}$.

The temperature-dependent XANES analysis of the Ge K-edge data showed, for the two samples, a gradual shift in the energy threshold $E_0$ during the amorphous-to-crystalline transition, suggesting a gap closure process. When comparing the amorphous and crystalline states of GeSe or $GeSe_{0.75}Te_{0.25}$, Ge and Se K-edges XANES spectra reveal differences in the local electronic structure and post-edge feature evolutions which are an indication of local and medium-range order structural modifications. Therefore, to access the quantitative information on the local-range order around the different chemical elements for the samples in the amorphous state, extended X-ray absorption fine structure (EXAFS) experiments should be performed.

**Funding:** This work was supported by the French Ministry of Higher Education, Scientific Research and Innovation, and the French National Center of Scientific Research (CNRS).

**Acknowledgments:** The room-temperature X-ray scattering experiments were done at the Platform of Analysis and Characterization (PAC) of the Pôle Chimie Balard in Montpellier, France. The *in-situ* temperature-dependent experiments (proposal ID 20210201) were performed at Synchrotron SOLEIL facility (France) on the DiffAbs beamline. C. Mocuta and P. Joly (DiffAbs beamline) are acknowledged for experimental and technical support during the experiments.